\documentstyle[12pt,amssymb,amsfonts,amsbsy]{article}
\begin{document}
\author{Ilja Schmelzer}
\thanks{Berlin, Germany}
\date{}

\title{General Ether Theory and Graviton Mass}

\maketitle

\begin{abstract}
\sloppypar
For negative cosmological constant $\Lambda$ we show the equivalence
of the Lagrangian of ``general ether theory'' with Logunov's
``relativistic theory of gravity'' with massive graviton and a variant
of GR with four non-standard scalar fields.

We consider the remaining differences between these theories.

\end{abstract}

\section{Introduction}

In \cite{Schmelzer}, a metric theory of gravity in a predefined
Newtonian framework with Galilean coordinates $T(x),X^i(x)$ with
Lagrange density

\[L = (R - \Lambda)\sqrt{-g} + L_{matter}(g_{\mu\nu},\psi^m)
    + \Xi g^{\mu\nu}\delta_{ij}X^i_{,\mu}X^j_{,\nu}\sqrt{-g}
    - \Upsilon g^{\mu\nu}T_{,\mu}T_{,\nu}\sqrt{-g}
\]

This theory allows a simple condensed matter interpretation.  This
condensed matter (``ether'') interpretation may be used to derive the
Lagrange density.  In this derivation, the constants $\Xi$, $\Upsilon$
and Einstein's cosmological constant $\Lambda$ remain unspecified,
even their signs.  We add here another simple hypothesis which
allows to specify the signs: we assume that there exists an
``undistorted reference state'' -- a solution with constant $\rho,
v^i, \sigma^{ij}$ -- and that this undistorted reference state is
stable.  This hypothesis fixes the signs as $\Xi>0$, $\Upsilon>0$.
Moreover, it requires $\Lambda<0$ for Einstein's cosmological
constant.

With these sign conventions, the Lagrange density may be transformed
in the preferred coordinates into

\[L = R\sqrt{-g} + L_{matter}(g_{\mu\nu},\psi^m)
    - m_g^2({1\over 2}\eta_{\mu\nu}g^{\mu\nu} - 1)\sqrt{-g}
\]

where $\eta_{\mu\nu}$ defines the vacuum solution and $m_g$ the mass
of the graviton in the vacuum state.  This was an unexpected result --
the Lagrangian looks like the usual GR Lagrangian with some additional
scalar fields $X^i(x)$, $T(x)$, and for $\Xi>0$, $\Upsilon>0$ they
simply lead to additional ``dark matter'' terms with pressure
$p=-{1\over3}\varepsilon$ resp. $p=\varepsilon$.  Therefore, I have
assumed that the graviton remains mass-less.

After this, it was reasonable to compare the theory with existing
theories with massive graviton.  This search has been successful, we
have found a theory with the same Lagrangian -- ``relativistic theory
of gravity'' developed by Logunov a.o. \cite{Logunov1}.

Thus, the Lagrangian has been derived independently based on
completely different motivation.  This is not strange, because the
harmonic condition -- the simplest and most beautiful coordinate
condition -- is used in above theories and is all what is necessary to
obtain the Lagrange formalism.

Some interesting properties of the theory in the limit of very small
$m_g\to 0$ are easy to understand -- we obtain an oscillating
universe, stable ``frozen stars'' of $\gtrapprox$ Schwarzschild size,
a bounce for the gravitational collapse.  Thus, it is a nice
regularization of GR.  Even for arbitrary small $m_g$ this solves
cosmological problems -- the horizon problem and the flatness problem
-- solved in standard cosmology with inflation (cf. \cite{Primack},
p.5,56).  Thus, we do not have to introduce inflation to solve these
problems.

Above theories have the same Lagrangian, but there are not only major
differences in the metaphysical interpretation, but also minor but
interesting differences in predictive power and differences in the
quantization concepts related with these theories.

\section{General Ether Theory}

The basic formula is the definition of the physical
metric $g^{\mu\nu}$ as a function of typical condensed matter
variables (density $\rho$, velocity $v^i$, stress tensor
$\sigma^{ij}$):

\begin{eqnarray} \label{gdef}
 \hat{g}^{00} = g^{00} \sqrt{-g} &=  &\rho \\
 \hat{g}^{i0} = g^{i0} \sqrt{-g} &=  &\rho v^i \\
 \hat{g}^{ij} = g^{ij} \sqrt{-g} &=  &\rho v^i v^j - \sigma^{ij}
\end{eqnarray}

This matter (the ``ether'') fulfils classical conservation laws:

\begin{eqnarray} \label{conservationlaw}
\partial_t \rho + \partial_i (\rho v^i) &= &0 \\
\partial_t (\rho v^j) + \partial_i(\rho v^i v^j - \sigma^{ij}) &= &0
\end{eqnarray}

Additional ``inner steps of freedom'' $\psi^m(x)$ are also allowed,
but no other, external matter.  Thus, there are no momentum exchange
terms for interaction with other matter.  The ``inner steps of
freedom'' of the ``ether'' are identified with usual matter fields.

In the metric variables, the conservation laws transform into the
harmonic equation for the Galilean coordinates:

\[\Box X^i = \Box T = 0 \]

If we search for a Lagrange density $L(g^{\mu\nu},\psi^m,X^i,T)$
which leads to these equations, we obtain the Lagrangian

\[L = L_{GR}(g_{\mu\nu},\psi^m)
    + \Xi g^{\mu\nu}\delta_{ij}X^i_{,\mu}X^j_{,\nu}\sqrt{-g}
    - \Upsilon g^{\mu\nu}T_{,\mu}T_{,\nu}\sqrt{-g}
\]

with unknown constants $\Xi$, $\Upsilon$ almost immediately: the
simplest way to obtain the harmonic equation for a field is the
standard scalar Lagrangian plus the requirement that the remaining
part does not depend on this field. But the requirement that the
remaining part does not depend on the preferred Galilean coordinates
is the requirement for the Lagrangian of GR.

This Lagrange formalism and the choice of independent variables seems
strange from point of view of classical condensed matter theory.  It
defines a promising analogy between condensed matter theory and
fundamental particle theory which is far away from being completely
understood.

In this derivation, the signs of the cosmological constants remain
unspecified.  They should be defined by observation.  Current
observation of the ``dark matter'' seems to favour $\Xi>0, \Lambda>0$.
For $\Upsilon>0$ we obtain interesting new effects -- especially, we
can solve the cosmological horizon problem without introducing
inflation theory.  Therefore, we tend to favour $\Upsilon>0, \Xi>0,
\Lambda>0$.

\subsection{GET with negative cosmological constant}

Nonetheless, another sign convention seems very interesting from
theoretical point of view: $\Upsilon>0, \Xi>0, \Lambda<0$.  In this
case, there exists a constant stable ``vacuum solution''. Indeed, if
there is no matter, the equation for the constant solution is:

\begin{eqnarray*}
ds^2 &=& a^2dt^2 - b^2(dx^2+dy^2+dz^2) \\
G^0_0 = 0  &=& -\Upsilon a^{-2} +3\Xi b^{-2}+\Lambda\\
G^1_1 = 0  &=& +\Upsilon a^{-2} + \Xi b^{-2}+\Lambda
\end{eqnarray*}

with $\Lambda = - 2 \Xi b^{-2} = - 2 \Upsilon a^{-2}$ as the only
solution.  In this situation,
it seems natural to renormalize the constants and to introduce the
vacuum state as $\eta^{\mu\nu}$ into the Lagrange density.

\[L = R\sqrt{-g} + L_{matter}(g_{\mu\nu},\psi^m)
    + \Lambda({1\over 2}\eta_{\alpha\beta}g^{\mu\nu}
              X^{\alpha}_{,\mu}X^{\beta}_{,\nu} - 1)\sqrt{-g}
\]

It is stable if we choose $\Lambda<0$.  Indeed, let's consider the
linearized equations for a small modification of the undistorted state
$g^{\mu\nu}(x)=\eta^{\mu\nu}+h^{\mu\nu}(x)$. We obtain

\[ {1\over 2}g^{kl}{\partial h^{ij}\over\partial x^k \partial x^l} =
-{\Lambda\over 2} h^{ij} + T^{ij} - {1\over 2}g^{ij}T \]

Thus, the theory becomes a theory with massive graviton with mass
$m_g=\sqrt{-\Lambda}$.  We obtain:

\[L = R\sqrt{-g} + L_{matter}(g_{\mu\nu},\psi^m)
    - m_g^2({1\over 2}\eta_{\alpha\beta}g^{\mu\nu}
              X^{\alpha}_{,\mu}X^{\beta}_{,\nu} - 1)\sqrt{-g}
\]

In the preferred coordinates, this Lagrange density is

\[L = R\sqrt{-g} + L_{matter}(g_{\mu\nu},\psi^m)
    - m_g^2({1\over 2}\eta_{\mu\nu}g^{\mu\nu} - 1)\sqrt{-g}
\]

\section{Comparison of GET with similar theories}

The Lagrange density for GET if $\Xi>0,\Upsilon>0,\Lambda<0$ is
equivalent to the Lagrange density in
\cite{Logunov3}, formulas (9),(10), for the ``relativistic theory of
gravity'' (RTG) with non-zero graviton mass.  On the other hand, if
$\Xi>0,\Upsilon<0$ we can formally obtain a similar Lagrangian if we
introduce ``clock fields'' $X^\mu(x)$ as scalar fields into GR (Kuchar
\cite{Kuchar} has considered such theories).  Thus, this Lagrangian
occurs with different motivation in three theories with completely
different metaphysics. Let's introduce the following notions:

 \begin{itemize}

 \item $\Lambda$DM -- a variant of general relativity with dark matter
fields $X^\mu(x)$ (no background):

\[L = (R-\Lambda)\sqrt{-g} + L_{matter}(g_{\mu\nu},\psi^m)
    - \Xi g^{\mu\nu}\eta_{\alpha\beta}X^\alpha_{,\mu}X^\beta_{,\nu}
\]

 \item RTG -- Logunov's relativistic theory of gravity with massive
graviton (Minkowski background):

\[L = L_{Rosen} + L_{matter}(g_{\mu\nu},\psi^m)
    - m_g^2({1\over 2}\eta_{\mu\nu}g^{\mu\nu}\sqrt{-g}
            - \sqrt{-g} - \sqrt{-\eta})
\]

There is an additional causality condition: the light cone of
$g_{\mu\nu}$ should be inside the light cone of
$\eta_{\mu\nu}$.

 \item GET -- the generalization of Lorentz ether theory to gravity
proposed by the author, for $\Upsilon>0, \Xi>0, \Lambda<0$ (Newtonian
background):

\[L = L_{GR}
    + \Xi g^{\mu\nu}\delta_{ij}X^i_{,\mu}X^j_{,\nu}\sqrt{-g}
    - \Upsilon g^{\mu\nu}T_{,\mu}T_{,\nu}\sqrt{-g}
\]

The causality condition in this theory is $g^{00}\sqrt{-g}>0$.

 \end{itemize}

Now, the common Lagrangian leads to common predictions which
distinguish these three theories from classical GR: stable ``frozen
stars'' near Schwarzschild size instead of black holes
\cite{Logunov2},\cite{Schmelzer}, with bounce after gravitational
collapse \cite{Logunov2}, a big bounce instead of a big bang
singularity \cite{Logunov1},\cite{Schmelzer} with an oscillating
universe \cite{Logunov1},\cite{Logunov3}.  Note that in an oscillating
universe there is no horizon problem, and we have a natural preference
for zero curvature.  That means, two of the problems used to justify
inflation theory (cf. \cite{Primack}) disappear.

These seem to be common effects of theories with massive graviton,
another way of introducing mass considered by Visser leads to similar
results about the behaviour near the horizon \cite{Visser}.  Even in
the limit $m_g\approx 0$ the qualitative differences remain.  Note
that once $\Upsilon>0$ $\Lambda$DM contains ``matter'' which violates
all energy conditions.  That's why GR theorems about black hole and
big bang singularities do not apply.

Let's now consider the differences.  First, we have a simple relation:
A solution of GET or RTG defines a solution of $\Lambda$DM.  In the
other direction, this is not correct.  The fields $X^\mu(x)$ of a
solution of $\Lambda$DM may not define a system of coordinates.  And
the solution of RTG possibly violates the condition $\rho(x)>0$.  That
means, if GET resp. RTG are true, $\Lambda$DM cannot be falsified. But
observing a solution of $\Lambda$DM where the four fields do not
define global coordinates -- for example, a solution with nontrivial
topology -- falsifies RTG and GET without falsifying $\Lambda$DM.
In this sense, RTG and GET have higher predictive power.

To compare GET and RTG, the causality condition seems to be important.
The causality condition of RTG is stronger than the causality
condition of GET.  Therefore, there may be GET-solutions which violate
RTG causality.  In this sense, the predictive power of RTG is higher.
In GET, there is also no restriction for the sign of $\Lambda, \Xi,
\Upsilon$, while RTG fixes these signs uniquely.

An interesting question about the physical meaning of the causality
conditions in RTG and GET is if there are solutions which do not
violate causality conditions for the initial values and what is their
physical meaning.  The answer for GET is that for solutions there
$\rho$ becomes zero the ``continuous large scale approximation''
described by GET fails, and we observe new physics -- atomic ether
theory.

Note that there are other important differences between GET and RTG.
GET is compatible with the EPR criterion of reality \cite{EPR} and
with realistic, deterministic, but non-local hidden variable theories
for QM like Bohmian mechanics \cite{Bohm}.  Therefore, despite the
fact that the three theories have equivalent Lagrangian formalism, and
therefore equivalent field equations, it is worth to distinguish them
as different physical theories.

These differences between the theories suggest essentially different
concepts for quantization.  In $\Lambda$DM we have the full beauty of
canonical GR quantization problems, especially the problem of time
\cite{Isham} and topological foam.  The harmonic condition may be
used, but is only a gauge condition \cite{Kuchar}.  Instead, in GET
and RTG it is a physical equation.  The conceptual problems of GR
quantization related with the absense of an absolute background are
not present.

In GET quantization we can use condensed matter analogies (see for
example Volovik \cite{Volovik}) as simple guiding principles.  This
suggests to use some ``atomic ether'' hypothesis which leads to an
explicit, physical regularization.  The ``ether hypothesis'' allows to
obtain a prediction about the cutoff length: $\rho(x) V_{crit} = 1$ in
appropriate units.  This prediction violates the relativistic
invariance of the Lagrangian and is therefore incompatible with RTG
symmetry.  It differs also from Planck length suggested by the
``Planck ether'' concept \cite{Jegerlehner}, \cite{Volovik}.


\begin{thebibliography}{99}

\bibitem{Aspect}
A. Aspect et.al., Phys. Rev.Lett. 49,
1801-1807, 1982

\bibitem{Bell}
J.S.~Bell, Speakable and unspeakable in quantum mechanics, Cambridge
University Press, Cambridge, 1987

\bibitem{Bohm}
D.~Bohm, Phys. Rev. 85, 166-193, 1952

\bibitem{EPR}
A.~Einstein, B.~Podolsky, N.~Rosen, Phys. Rev. 47, 777-780,
1935

\bibitem{Fock}
V.~Fock, Theorie von Raum, Zeit und Gravitation, Akademie-Verlag
Berlin, 1960

\bibitem{Isham}
C.~Isham, gr-qc/9210011

\bibitem{Jegerlehner}
F.~Jegerlehner, hep-th/9803021

\bibitem{Kuchar}
K.V.~Kuchar, C.G.~Torre, Physical
Review D, vol. 44, nr. 10, pp. 3116-3123, 1991

\bibitem{Logunov1}

A.A.~Logunov, M.A.Mestvirishvili, Yu.V.Chugreev, Graviton mass and
evolution of a Friedman universe, Theor.Math.Phys. 74, 1-20 (1988)

\bibitem{Logunov2}

A.A. Vlasov, A.A. Logunov, Bouncing from the Schwarzschild sphere in
the relativistic theory of gravity with nonzero graviton mass,
Theor.Math.Phys. 78, 229-233 (1989)

\bibitem{Logunov3}
S.S. Gershtein, A.A. Logunov, M.A.Mestvirishvili, The upper limit on
the graviton mass, hep-th/9711147

\bibitem{Primack}
J.R.~Primack, astro-ph/9707285

\bibitem{Schmelzer}
I.~Schmelzer, gr-qc/9811033

\bibitem{Visser}
M. Visser, Mass for the graviton, gr-qc/9705051

\bibitem{Volovik}
G.E.~Volovik, Induced gravity in superfluid $^3He$, cond-mat/9806010

\end{thebibliography}
\end{document}